# Considerations on Gravitational Effects Stimulated by Gravitational Fields via Classical Field Theory

Helmut Kling *

(Dated: October 7, 2013)

**Abstract**

A mass distribution is analyzed in terms of classical gravitational field theory. Newton's law of gravitation is consistently applied on the assumption that the equivalence of energy and mass according to Einstein's theory of relativity is valid for gravitational fields as well. Different from standard approaches the gravitational field, via its associated field energy, is handled as a source of gravitation by itself. Starting from these principles a gravitational self-shielding phenomenon is derived as a common characteristic of all mass/energy distributions. Moreover, it is demonstrated that even in the absence of any stimulating mass the existence of independent static gravitational fields is fully consistent with Newton's law of gravitation as long as the equivalence of energy and mass is respected. From a distance such gravitational fields appear as negative masses.

## 1. Introduction

When analyzing the gravitation of a mass distribution by means of classical field theory Newton's law of gravitation is applied to the stimulating mass distribution. According to Einstein's theory of special relativity energy and inertial mass are equivalent. Einstein's theory of general relativity extended this principle to the gravitational mass. Consequently, the term stimulating mass has not only to include classical masses but any energy distribution. In classical field theory, however, the associated gravitational field is usually not seen as a source of gravitation although it represents an energy distribution.

In this study a mass distribution is analyzed in terms of classical gravitational field theory in which the gravitational field energy is handled as a mass, which is acting as a source of gravitation by itself. In that sense equivalence of energy and mass is applied more consequently as is usually done. Based on these principles a field equation for static gravitational fields is developed and applied to a spherical mass distribution and also to the special case where no stimulating mass is present at all, i.e. to vacuum.

Beyond the described equivalence of energy and gravitational mass further aspects of Einstein's theory of general relativity, its formalism including Einstein's field equations are not in the scope of this work.

## 2. Gravitational Field Strength and Energy Density

Analysis shall be started from Newton's law of gravitation, which gives the following force to a particle of mass $m$ at the position $x$, which is exposed to the gravitation of a second mass $m_2$ at the position $x_2$

$$\boldsymbol{F}_g = -\,G\,m\,m_2\,\frac{\boldsymbol{x}-\boldsymbol{x}_2}{|\boldsymbol{x}-\boldsymbol{x}_2|^3} \quad , \tag{1}$$

where G is the gravitational constant. We get

$$\boldsymbol{F}_g(\boldsymbol{x}) = -\,G\,m\,\int \rho(\boldsymbol{x}')\frac{\boldsymbol{x}-\boldsymbol{x}'}{|\boldsymbol{x}-\boldsymbol{x}'|^3}\,d^3x' \quad , \tag{2}$$

if the origin of the gravitational force is not a single point mass but a mass distribution represented by $\rho(x)$. This mass distribution is associated with a gravitational field described by its field strength

$$\boldsymbol{E}_g(\boldsymbol{x}) = -\,G\,\int \rho(\boldsymbol{x}')\frac{\boldsymbol{x}-\boldsymbol{x}'}{|\boldsymbol{x}-\boldsymbol{x}'|^3}\,d^3x' \tag{3}$$

so that $\boldsymbol{F}_g = m\,\boldsymbol{E}_g$. Using $\nabla(1/|\boldsymbol{x}-\boldsymbol{x}'|) = -(\boldsymbol{x}-\boldsymbol{x}')/|\boldsymbol{x}-\boldsymbol{x}'|^3$ leads to

$$\boldsymbol{E}_g(\boldsymbol{x}) = G\,\nabla \int \frac{\rho(\boldsymbol{x}')}{|\boldsymbol{x}-\boldsymbol{x}'|}\,d^3x' \quad . \tag{4}$$

The gravitational field carries energy at a density of

$$u_g = -\frac{1}{8\pi G}\,E_g^2 \quad . \tag{5}$$

This expression can easily be obtained in full analogy to the corresponding electrostatic case. Detailed evaluations of the electrical case are given in any textbook on electrodynamics such as [1] or [2]. However, even if well known [3] it has to be noted that different from electrostatics gravitational field energy is always negative, which is a consequence of Newton's law of gravitation when respecting the law of the conservation of energy due to the opposite sign compared to Coulomb's law.

### 3. Evaluation of the Gravitational Field Strength

Every mass/energy distribution is accompanied by a gravitational field carrying an energy density, which is, according to Einstein's correlation E = mc² (E energy, m mass, c vacuum speed of light), equivalent to a mass distribution. As Visser pointed out in his paper "A Classical Model for the Electron" [4][1], it is logic and consistent to assume that the gravitational field acts again as a source of gravitation and the overall mass density to be taken into account is given by $\rho_{tot} = \rho + \rho_{field} = \rho + u_g/c^2$, where $\rho$ represents the density of the stimulating mass as introduced in section 1 and $\rho_{field}$ the equivalent mass density resulting from the gravitational field energy given by equation (5). It is obvious that this field-induced contribution, due to the property of the gravitational field energy to be negative, is equivalent to a negative mass distribution and acts opposite to the stimulating mass distribution: It weakens its gravitational effect. The gravitational field strength can therefore be expressed as

$$\boldsymbol{E}_g(\boldsymbol{x}) = G \nabla \int \left[\rho(\boldsymbol{x}') + \frac{u_g}{c^2}\right] \frac{d^3 x'}{|\boldsymbol{x} - \boldsymbol{x}'|} \tag{6}$$

resulting in the following field equation on $\boldsymbol{E}_g$

$$\boldsymbol{E}_g(\boldsymbol{x}) = G \nabla \int \left[\rho(\boldsymbol{x}') - \frac{1}{8\pi G c^2} E_g^2(\boldsymbol{x}')\right] \frac{d^3 x'}{|\boldsymbol{x} - \boldsymbol{x}'|} \quad . \tag{7}$$

Let us now concentrate our analysis on a spherical mass distribution having a radius a and a mass M, whose center is located at the origin of our coordinate system. Spherical symmetry implies that $\boldsymbol{E}_g(\boldsymbol{x})$ is not dependent on polar angles and always pointing into the direction of $\boldsymbol{e}_r$, which is the radial unit vector. When orientating our coordinate system in such a way that $\boldsymbol{x}$ points along the z axis, i.e. $\vartheta = 0$, equation (7) can be rewritten as

$$E_g(r) = - G \frac{M}{r^2} - \frac{1}{8\pi c^2} \nabla_r \int_a^\infty E_g^2(r') r'^2 dr' \int_0^{2\pi} d\varphi' \int_0^\pi \frac{\sin \vartheta' \, d\vartheta'}{\sqrt{r^2 - 2 r r' \cos \vartheta' + r'^2}} \tag{8}$$

for r ≥ a, where r =|**x**|, $\varphi$ and $\vartheta$ spherical coordinates and $\boldsymbol{E}_g(\boldsymbol{x}) = E_g(r) \, \boldsymbol{e}_r$.

---

[1] In [4] M. Visser considers the gravitational field as well as the electrical field as a source of gravitation and applies this principle to his classical model for the electron, which assumes the electron to be a point particle. Other arrangements are not investigated in this paper.

Performing integration on $\varphi$ and using

$$\int_0^\pi \left(\sin\vartheta'/\sqrt{r^2 - 2rr'\cos\vartheta' + r'^2}\right) d\vartheta' = (r + r' - |r - r'|)/rr' \tag{9}$$

we end up at

$$E_g(r) = -G\frac{M}{r^2} - \frac{1}{2c^2}\nabla_r\left[\frac{1}{r}\int_a^r E_g^2(r')r'^2 dr' + \int_r^\infty E_g^2(r')r' dr'\right]$$

$$= -G\frac{M}{r^2} + \frac{1}{2c^2}\frac{1}{r^2}\int_a^r E_g^2(r')r'^2 dr' \quad. \tag{10}$$

Equation (10) reflects the well known fact that the gravitational field at a distance r from the center of a spherical mass distribution is equal to the field of a point mass located at the center whose magnitude is identical to the total mass enclosed by the sphere of radius r. In order to simplify the evaluation of this integral equation we substitute $\kappa(r) = -r^2 E_g(r)/GM$ assuming M>0, which leads to

$$\kappa(r) = 1 - R_g\int_a^r \kappa^2(r')/r'^2 dr' \quad, \tag{11}$$

where $R_g = GM/2c^2$ is one fourth of the Schwarzschild radius [5]. Differentiation yields

$$\frac{d\kappa(r)}{dr} = -R_g\frac{\kappa^2(r)}{r^2} \quad. \tag{12}$$

Separation of the variables $\kappa$ and r results in

$$\frac{d\kappa(r)}{\kappa^2} = -R_g\frac{dr}{r^2} \quad. \tag{13}$$

Integration then leads to

$$\kappa(r) = \frac{1}{d - R_g/r} \quad, \tag{14}$$

where d is an integration constant. It can easily be calculated by turning back to equation (11). Finally we get

$$\kappa(r) = \frac{1}{1 + \frac{R_g}{a} - \frac{R_g}{r}} \tag{15}$$

and

$$E_g(r) = -\frac{1}{1 + \frac{R_g}{a} - \frac{R_g}{r}} G \frac{M}{r^2} \quad . \tag{16}$$

Evidently $\kappa(r)$ describes the influence of the gravitational field itself. Figure 1 shows **κ** versus r/a for different $R_g$/a. Obviously at the surface of the sphere no self-contribution of the

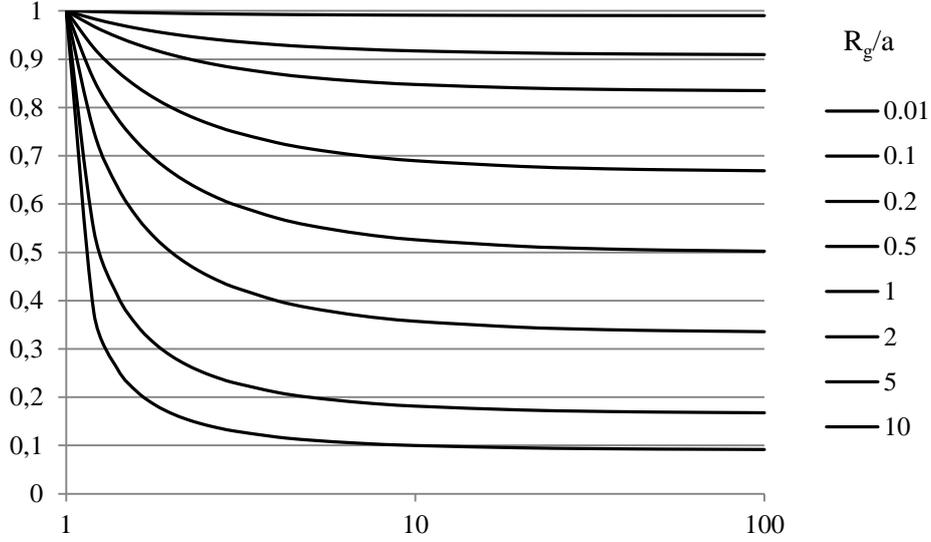

Figure 1: **κ** versus r/a for different $R_g$/a gravitational field is at work, which is confirmed by equation (15), i.e. $\kappa(r = a) = 1$. For $R_g >$ a another interesting reference point is r = $R_g$. In order to achieve a given gravitational field strength at this point according to equation (16) a mass is needed which is a factor $R_g$/a bigger than it would be expected when ignoring the contribution of the gravitational field itself. Let us now investigate the resulting behavior at long distances from the sphere, in particular for r $\gg$ $R_g$. Expansion of $E_g(r)$ in powers of $R_g/r$ yields

$$E_g(r) = -\frac{1}{1 + \frac{R_g}{a}} \left[ 1 + \frac{1}{1 + \frac{R_g}{a}} \frac{R_g}{r} + \left( \frac{1}{1 + \frac{R_g}{a}} \frac{R_g}{r} \right)^2 + \ldots \right] G \frac{M}{r^2} \quad . \tag{15}$$

Restriction to the first term leads to

$$E_g(r) = -\frac{1}{1 + \frac{R_g}{a}} G \frac{M}{r^2} \quad . \tag{16}$$

When observing the gravitational interaction at a long distance the sphere will appear as an object having a mass $M_{obs}$ according to classical theory of gravitation, i.e. when ignoring the contribution of the gravitational field itself, which is

$$M_{obs} = \frac{M}{1 + \frac{R_g}{a}} = (1 - \sigma) M \quad , \tag{17}$$

where $R_g = GM/2c^2$ and $\sigma = 1 - 1/(1 + R_g/a)$. $\sigma$ is a self-shielding coefficient measuring, which amount of the gravitational field strength is shielded by the gravitational field itself. $\sigma$ is governed by $R_g/a$, which effectively means by the ratio $M/a$ of the stimulating spherical mass distribution. Table 1 gives self-shielding coefficients for selected objects. It is not a surprise that, fully in line with all observations, the self-shielding effects of conventional objects such as moon, earth and sun are too small to be visible. However, when calculating the self-shielding of a neutron star according to [9] we end up at $\sigma$ = 13%, which is not negligible any more. Table 2 shows self-shielding coefficients for different categories of black holes resulting in remarkable self-shielding coefficients. However, we have to be aware that any information on the physical radius of a black hole has to be handled very cautiously if it is deduced from a gravitational effect. An exciting question is now how strong this self-shielding effect could be. Is it imaginable that some kind of matter would exist, which is shielding its gravitation to a percentage close to 100% so that it would be difficult to observe its gravitational field at all? In table 3 a series of objects is modeled in order to approach this question. In order to evaluate, which mass/energy densities would be needed starting from a realistic micro black hole according to current knowledge models 1 – 3 are investigated when increasing the volume mass density in steps of a factor 1000. This procedure is repeated on models 1a – 3a for super massive black holes. In order to make sure not to leave any reasonable range of physics the associated volume mass density is given in this table as well and for reference the relevant parameters of the fictive Planck particle [10] are listed. The last line in table 3 gives the parameters of a fictive stellar black hole having on the one hand 10 times of the mass of the sun and on the other hand the typical volume density of a micro black hole. In this case the self-shielding coefficient would be >99.999%! The conclusion is that a high mass density is needed in order to achieve gravitational self-shielding close to 100%. Anyway, it seems imaginable that some kind of highly compact matter exists whose gravitational interaction at long distances is very weak.

| Object | Radius/m | M/kg | $R_g/a$ | $\sigma$ |
|---|---|---|---|---|
| Moon [6] | $1.74 \times 10^6$ | $7.35 \times 10^{22}$ | $1.57 \times 10^{-11}$ | $1.57 \times 10^{-11}$ |
| Earth[7] | $6.37 \times 10^6$ | $5.97 \times 10^{24}$ | $3.48 \times 10^{-10}$ | $3.48 \times 10^{-10}$ |
| Sun[8] | $6.96 \times 10^8$ | $1.99 \times 10^{30}$ | $1.06 \times 10^{-6}$ | $1.06 \times 10^{-6}$ |
| Neutron star [9] | $12 \times 10^3$ | $2.5\ M_{Sun}$ | 0.15 | 13% |

Table 1: Self-shielding coefficients $\sigma$ for selected objects

| Object | Radius | M/kg | $R_g/a$ | $\sigma$ |
|---|---|---|---|---|
| Micro Black Hole | 0.1 mm | $M_{Moon}$ | 0.27 | 21% |
| Stellar Black Hole | 30 km | $10\ M_{Sun}$ | 0.25 | 20% |
| Intermediate Mass Black Hole | 1000 km | $1000\ M_{Sun}$ | 0.74 | 42% |
| Super massive Black Hole | 10 au | $10^9\ M_{Sun}$ | 0.49 | 33% |

Table 2: Self-shielding coefficients $\sigma$ for black holes (1 au = $1.496\ \text{x}10^{11}$ m)

| Object | Radius | M | $R_g/a$ | $\sigma$ | Mass Density $kg/m^3$ |
|---|---|---|---|---|---|
| Planck particle [10] | $5.73\ \text{x}\ 10^{-35}\ m$ | $3.86\ \text{x}\ 10^{-8}\ kg$ | 0.25 | 20% | $4.9\ \text{x}\ 10^{94}$ |
| | | | | | |
| Micro Black Hole | 0.1 mm | $M_{Moon}$ | 0.27 | 21% | $1.75\ \text{x}\ 10^{34}$ |
| Model 1 | 0.01 mm | $M_{Moon}$ | 2.73 | 73% | $1.75\ \text{x}\ 10^{37}$ |
| Model 2 | 0.001 mm | $M_{Moon}$ | 27.3 | 96% | $1.75\ \text{x}\ 10^{40}$ |
| Model 3 | 0.0001 mm | $M_{Moon}$ | 273 | 99.6% | $1.75\ \text{x}\ 10^{43}$ |
| Super massive Black Hole | 10 au | $10^9\ M_{Sun}$ | 0.49 | 33% | $1.42\ \text{x}\ 10^2$ |
| Model 1a | 1 au | $10^9\ M_{Sun}$ | 4.94 | 83% | $1.42\ \text{x}\ 10^5$ |
| Model 2a | 0.1 au | $10^9\ M_{Sun}$ | 49.4 | 98% | $1.42\ \text{x}\ 10^8$ |
| Model 3a | 0.01 au | $10^9\ M_{Sun}$ | 494 | 99.8% | $1.42\ \text{x}\ 10^{11}$ |
| Stellar Black Hole Model 4 | 0.065 m | $10\ M_{Sun}$ | $10^5$ | >99.999% | $1.75\ \text{x}\ 10^{34}$ |

Table 3: Self-shielding coefficients σ for fictive black holes (models)

## *4.* Vacuum Solution

In the previous section an interesting effect has been derived which is relevant mainly for big, high density masses. In this section we will examine the other extreme, where no conventional mass is available at all, i.e. in a (classical) vacuum. Motivation for this investigation is arising from the conclusion in section 3 that the gravitational field acts as a source of gravitation by itself. The interesting question is now if there's a non-trivial solution to our field equation, a solution different from $\boldsymbol{E}_g(\boldsymbol{x}) = 0$ in the absence of any conventional mass, i.e. if $\rho(\boldsymbol{x}) = 0$?

Starting from equation (7) we get for $\rho(\boldsymbol{x}) = 0$

$$\boldsymbol{E}_g(\boldsymbol{x}) = -\frac{1}{8\pi c^2} \nabla \int_0^\infty \boldsymbol{E}_g^2(\boldsymbol{x}') \frac{d^3 x'}{|\boldsymbol{x} - \boldsymbol{x}'|} \;. \tag{18}$$

In order to strive for a simple solution let us again restrict our investigation to a spherically symmetric case, where $\boldsymbol{E}_g(\boldsymbol{x})$ is not dependent on polar angles and always pointing into the direction of $\boldsymbol{e}_r$, which is the unit vector in spherical coordinates. In this case we end up, in full agreement with equation (10), at

$$E_g(r) = \frac{1}{2c^2} \frac{1}{r^2} \int_0^r E_g^2(r') r'^2 dr' \;, \tag{19}$$

where again $\boldsymbol{E}_g(\boldsymbol{x}) = E_g(r)\, \boldsymbol{e}_r$ . Let us now solve this equation in the same as way we did in the previous section. Substituting $\zeta(r) = r^2 E_g(r)$ leads to

$$\zeta(r) = \frac{1}{2c^2} \int_0^r \zeta^2(r')/r'^2 dr' \;. \tag{20}$$

Differentiation yields

$$\frac{d\zeta(r)}{dr} = \frac{1}{2c^2} \frac{\zeta^2(r)}{r^2} \quad . \tag{21}$$

Separation of the variables $\zeta$ and r results in

$$\frac{d\zeta(r)}{\zeta^2} = \frac{1}{2c^2} \frac{dr}{r^2} \quad . \tag{22}$$

Integration leads to

$$-\frac{1}{\zeta} = -\frac{1}{2c^2} \frac{1}{r} + \frac{1}{Gd} \quad , \tag{23}$$

where d is a - properly scaled - integration constant when G again means the gravitational constant. For $\zeta(r)$ we find

$$\zeta(r) = -\frac{Gd}{1 - r_g/r} \quad , \tag{24}$$

where $r_g = Gd/2c^2$. Insertion into equation (20) reveals that this expression for $\zeta(r)$ is a solution for every d, $|d| < \infty$. Finally we get

$$E_g(r) = -\frac{1}{1 - r_g/r} \frac{Gd}{r^2} \tag{25}$$

for the gravitational field strength. For $r \gg |r_g|$ Equation (25) becomes

$$E_g(r) = -\frac{Gd}{r^2} \quad , \tag{26}$$

which is the gravitational field of a mass d. So the integration constant d turns out to be an effective mass stimulated by the gravitational field. Consequently let us write $d = M_{\text{field}}$. Now we have to remember that the investigated setup does not include any conventional mass and consists of gravitational field energy only. As gravitational field energy is always negative it is obvious that according to E=mc² also the related mass has to be negative and $r_g$ will be negative as well. This furthermore makes sure that we will not run into a problem with equation (25) at r = |$r_g$|. Let us now investigate $E_g$ also in the near field range, i.e. for $r \ll |r_g|$. Using $d = M_{\text{field}} = -|M_{\text{field}}|$ and $r_g = -|r_g|$ equation (25) can be rewritten as

$$E_g(r) = \frac{1}{1 + |r_g|/r} \frac{G|M_{field}|}{r^2}$$

$$= \frac{1}{r/|r_g| + 1} \frac{2c^2}{r} < \frac{2c^2}{r} \quad , \tag{27}$$

leading to

$$E_g(r) = \frac{2c^2}{r} \tag{28}$$

for r ≪ |r_g|. Obviously $2c^2/r$ is constituting an upper limit for $E_g(r)$ for all distances r. In the center of the gravitational field distribution $E_g(r)$ is approaching this limiting curve. In this range $E_g(r)$ is not dependent on the magnitude of $M_{field}$ anymore and just determined by the velocity of light, i.e. by a fundamental physical constant. At a distance $|r_g|$ from the center $E_g(r)$ is dropping to 50% of $2c^2/r$. So $M_{field}$ via $r_g$ is defining the range of influence of this gravitational field distribution. The maximum field strength, however, is not influenced by $M_{field}$ anymore. Furthermore for r ≪ |r_g| the classical $1/r^2$ behavior has been replaced by a 1/r dependency. Even if $E_g(r)$ still diverges for r → 0 the overall energy stored in this gravitational field

$$U_g = -\frac{1}{8\pi G} \int E_g^2(\boldsymbol{x'}) \, d^3x'$$

$$= -\frac{G|M_{field}|^2}{2} \int_0^\infty \frac{1}{\left(1 + |r_g|/r\right)^2} \frac{dr}{r^2}$$

$$= -|M_{field}|c^2 \int_0^\infty \frac{1}{\left(1 + |r_g|/r\right)^2} \frac{|r_g| \, dr}{r^2}$$

$$= M_{field} \, c^2 < 0 \tag{29}$$

is finite and fully consistent with our interpretation of the integration constant d. Evidently this gravitational energy distribution, according to equation (27), behaves in its far field as a negative mass. Its overall energy content is limited and meets the well known relation between energy and mass $E=mc^2$.

The conclusion from this analysis is that there is a non-trivial static solution of the gravitational field equation describing a spherical gravitational field distribution even if no stimulating mass is present at all. It is not within the scope of this study to investigate if or how such a gravitational field could be generated. However, it is interesting enough to realize that such a gravitational field distribution is theoretically possible according to Newton's law of gravitation when consistently respecting equivalence of energy and mass. Such a field distribution behaves as a negative mass, i.e. it is repelling all classical masses and attracting other energy distributions of the same gravitational nature. There are no restrictions on the magnitude of such a field distribution. If such a gravitational field distribution representing a big equivalent mass was located between two galaxies it would obviously push both galaxies away from each other and cause an expansion of this arrangement. When observing such a gravitational energy distribution from a far distance it would appear as a dark area as it is repelling not only masses but also light.

## *5.* Conclusions

It has been shown that consistent application of classical field theory results in the prediction of interesting, from a classical perspective unexpected phenomena if the equivalence of energy and mass is accepted for the gravitational field as well. These effects are introduced by the fundamental assumption that gravitational interaction is associated with a gravitational field having an energy content, which is obeying the law of the conservation of energy and which is again equivalent to a mass distribution stimulating gravitation by itself. By this means a gravitational self-shielding mechanism is derived as a common property of all mass/energy distributions. Weakness of gravitational interaction makes sure that these gravitational self-shielding effects are very small for conventional objects such as earth or sun. Significant effects, however, are indicated for high density objects such as neutron stars or black holes. Moreover, it has been demonstrated that these basic principles are compatible with the existence of independent static gravitational fields even in the absence of any stimulating conventional mass.

## *6.* Acknowledgement

Thanks are due to Professor W. Hüttner, University of Ulm for valuable comments.

**References**

*   helmutkling@vodafone.de;

    permanent address: Junginger Strasse 24, 89275 Elchingen, Germany